\documentstyle[prb,aps]{revtex}

\begin{document}
\draft
\title{A potential inversion study of liquid CuBr}

\author{J. Neuefeind} 
\address{Hamburger Synchrotronstrahlungslabor HASYLAB at 
Deutsches Elektronensynchrotron DESY, 
Notkestr 85, D-22603 Hamburg Germany}
\date{\today}
\maketitle
\begin{abstract}
Hard electromagnetic radiation diffraction experiments on copper(I)-bromide 
melt are presented. Combining the result of this experiment with 
earlier neutron 
diffraction experiments the partial pair distribution functions of CuBr 
have been determined. The differing results for two of these functions
obtained recently by various techniques are discussed. A potential inversion 
scheme has been applied to generate three dimensional structures from the 
partial pair distribution function. The angular correlations between near 
neighbor atoms have been determined. These show characteristic differences 
with the glass-forming ZnCl$_2$ melt. While in ZnCl$_2$ melt the 
angle joining adjacent ZnCl$_4$ tetrahedra has been found 
- as in silica glass - well defined and the $+-+$ (ZnClZn) angle 
distribution peaked the corresponding distribution function in CuBr is 
broad. This is probably a key to understand, why ZnCl$_2$ but not CuBr melt 
can be easily supercooled into a glassy state.
\end{abstract}


\section{Introduction}
Copper-(I)-bromide, like several other copper and silver halides,
attains a rather large ionic conductivity before melting,
and belongs to a class of ionic systems called fast-ion conductors.
A representative of this class of substances, the CuCl, belongs to the first 
substances investigated by the neutron diffraction with isotopic substitution 
(NDIS) technique \cite{PM71}. In this early work, a featureless CuCu partial 
pair distribution function (PPDF) $g_{CuCu}$ has been found, 
and this structural 
detail attracted much attention - including a redetermination 
and confirmation of the PPDF \cite{eisen} --
as it provides an easy explanation for the fast-ion conductivity: The 
small copper ions move more or less gas-like through the network of 
bromine atoms. According to a hypotheses of Ginoza {\it et al} \cite{Gin87}
a structureless cation-cation PPDF should be common to all fast 
(cat)ion conductors.

The structure of liquid CuBr has been 
investigated by Allen {\it et al.} \cite{All92}
by NDIS making use of the different neutron coherent scattering lengths 
of\ $^{65}$Cu(6.72 $\cdot$\,10$^{-15}$\,m) 
and $^{63}$Cu(11.09 $\cdot$\,10$^{-15}$\,m) \cite{Koe91} to determine 
the PPDF. Again, a flat, gas-like
$g_{CuCu}$ was found in this study. The 
structure determination by Allen {\it et al.} has been challenged 
by DiCicco {\it et al.} \cite{DiC97}, especially concerning the 
$g_{CuBr}$ PPDF. 
This group used extended
x-ray absorption fine structure (EXAFS)  spectroscopy at 
both the copper and the bromine edge to investigate the short range 
order in CuBr(l), and found the first shell bromine coordination around 
copper significantly sharper than determined with neutron diffraction. 
This is taken by DiCicco {\it et al.} as an indication that apart from
inter-ionic repulsion there is a significant contribution of
covalency to the inter-atomic potential. Further, reverse Monte Carlo (RMC) 
simulations gave evidence \cite{Pus98} that the flat $g_{CuCu}$ is likely 
to be an artifact of the maximum entropy analysis used in ref.~\onlinecite{All92}.

This paper will address the question about the sharpness of the 
first CuBr coordination  shell. A set of partial structure factors (PSF)  is 
determined consistent with all available diffraction information.  
Using the potential inversion scheme of Levesque, Weis, Reatto 
\cite{LWR} an empirical two body potential is developed. This allows to check
the existence of a three-dimensional arrangement of atoms consistent with 
this set of PSF and  to interprete  the one-dimensional
pair distribution functions in terms of three-dimensional 
structures. The structure of both
CuBr and ZnCl$_2$ melt is characterized by tetrahedral structural units - 
Zn(II) is isoelectronic with Cu(I).
The angular correlation between neighboring atoms is hence discussed 
comparing these two systems.

\section{Experimental}
Liquid diffraction experiments have been carried out at the high energy
beam-line BW5 at the DORIS-III storage ring at HASYLAB in Hamburg, in the 
set-up for liquid and amorphous substances \cite{Bou98}. 
The photon energy  was set to 121.0 keV. The covered momentum transfer 
range 
was 0.43 \AA$^{-1} < Q < $ 26.5 \AA$^{-1}$, with 
$Q=4\pi/\lambda \sin(\theta)$,
$\lambda$ is the wave-length and $\theta$ half the scattering 
angle.

The sample (Sigma Chemical Co., quoted purity $>$99.8\%) 
was sealed under vacuum in a quartz-glass tube of 5 mm inner
diameter. The sample was heated to 803 K in a standard neutron diffraction 
furnace \cite{YB}. This furnace -- capable to reach a maximum temperature of 
2100K -- contains several niobium heat shields the inner most with a 
radius of 50\, mm. As in ref.~\onlinecite{Neu98} a restrictive collimation has been used to 
reduce the background scattering contribution of these niobium shields.
Fig.~\ref{abb-roh-cubr} shows the scattering contribution of background 
and sample.
The main  contribution to the background is the scattering from 
the silica container. At low momentum transfers, some powder lines 
from the niobium shields are remaining in spite of the collimation system. 
The data points at the 
positions of these niobium lines are excluded from the further analysis.
The usual corrections for detector dead-time, absorption and 
polarization are made and the intensities are converted into a differential 
cross section normalizing to the known cross-section (the sum of elastic
self scattering and Compton cross section) at large momentum 
transfers. The data correction procedure is described in more detail in 
\cite{MolPhys}.
\section{Theoretical summary}

The differential cross section of a liquid in a neutron or electromagnetic 
radiation scattering experiment can be expressed in terms of a total 
scattering function:
\begin{eqnarray}
S^{(n)}(Q)&=&\frac{\left(\frac{d\sigma}{d\Omega}\right)^{(n)}-
\sum_{i}^{N_{uc}} \nu_i b_i^2}{(\sum_{i}^{N_{uc}} \nu_i b_i)^2}+1 
\label{eq-sq-n}\\
S^{(x)}(Q)=i(Q)+1&=&\frac{\left(\frac{d\sigma}{d\Omega}\right)^{(x)}/
\sigma_{el}-\sum_{i}^{N_{uc}} \nu_i f_i^2}{(\sum_{i}^{N_{uc}} \nu_i f_i)^2}
+1\label{eq-sq-x}
\end{eqnarray}
where $\left(\frac{d\sigma}{d\Omega}\right)$ is the coherent 
differential cross section, $Q=4\pi/\lambda \sin(\theta)$,$\lambda$ the 
wavelength of the radiation and $\theta$ the diffraction angle, 
$b_i$ the coherent scattering lengths \cite{Koe91}, $f_i$ the X-ray form 
factors in the independent atom approximation \cite{Hub75}, 
$\sigma_{el}$ the scattering cross section of the free electron, $\nu_i$ the 
stoichiometric coefficient of the atom $i$, and where the sums are extending 
over the number of distinct atoms $N_{uc}$ in the unit of composition, 
CuBr. 
These total structure functions are composed of partial structure factors 
$s_{ij}$:
\begin{equation}
S^{(n/x)}= \sum_{ij} w_{ij}(Q) s_{ij} \label{eq-PSF}
 \end{equation} 
where
\begin{equation}
w_{ij}(Q) = \frac{\nu_i \nu_j f_i(Q) f_j(Q)}{(\sum \nu_i f_i(Q)^2}
\label{eq-wij}
\end{equation} 
and 
\begin{equation}
\sum_{ij} w_{ij}(Q)=1\, .
\end{equation}
For the neutron case $f(Q)$ has to be replaced by $b$ and the 
weighting factors  become independent of $Q$.

The total pair distribution functions are related to the structure factors 
of Eqs. \ref{eq-sq-n} and \ref{eq-sq-x} via: 
\begin{equation}
r \cdot (^{(n/x)}g-1) = \frac{1}{2\pi^2\rho_{uc}} 
\int Q \cdot (^{(n/x)}S(Q)-1)  
\sin(Qr) dQ \label{eq-ft}
\end{equation}
with $\rho_{uc}$ the density per unit of composition. 
In this work a value of $\rho_{uc}=0.0175$\,\AA$^{-1}$ for CuBr at 
800\, K has been used to keep consistency with the work of 
Allen and Howe \cite{All92}, although an experimental value of  
$\rho_{uc}=0.0181$\,\AA$^{-1}$ has been reported later by Saito {\it et al} 
\cite{Sai97}.
Likewise, the partial structure factors are related to the 
partial pair distribution functions (PPDF) via :
\begin{equation}
r \cdot (g_{ij}-1) = \frac{1}{2\pi^2\rho_{uc}} 
\int Q \cdot (s_{ij}-1)  
\sin(Qr) dQ \label{eq-ft-n}
\end{equation}
and, hence, the total pair distribution:
\begin{equation}
^{n/x}g(r) = \sum_{ij} \mbox{FT}[w_{ij}(Q)] \otimes g_{ij} \label{eq-conv}
\end{equation}
where FT is the Fourier sine transformation and $\otimes$ 
the convolution operation. The convolution reduces to a simple product 
in the case of neutron scattering. 

An alternative definition of the neutron total structure factor has been 
used by Allen and Howe \cite{All92}:
\begin{equation}
F(Q)=c_a^2 b_a^2 [S_{aa}(Q)-1] + 2 c_a c_b b_a b_b [S_{ab}(Q)-1] + c_b^2 b_b^2 [S_{bb}(Q)-1] \label{eq-f}
\end{equation}
where $c_a$, $c_b$ are the concentration of the atomic species, 
the $S_{aa}$, $S_{bb}$, $S_{bb}$ are partial structure factors differing from
the $s_{ij}$ in Eq. \ref{eq-PSF} by a factor of $\sum \nu$. It is noted, 
that in this definition the total neutron structure factor has the 
dimension of a cross section, while $S^{(n)}(Q)$ in Eq. \ref{eq-sq-n} 
is dimensionless.

In order to generate three dimensional structures from the pair distribution 
functions, the potential inversion scheme of Levesque, 
Weiss and Reatto \cite{LWR} (LWR-scheme) has been applied. The idea of 
this method is based on the equation:
\begin{equation}
\label{eq-hd}
g(r)= \exp \left[ \frac{-v(r)}{kT}  + g(r) - 1 - c(r) + B(r,v)\right] 
\label{eq-B}
\end{equation}
relating the pair distribution function and the pair potential, where
$v(r)$ is the pair potential, $c(r)$ the direct correlation function and 
B(r,v) the bridge function. Starting with a crude guess 
of the bridge function,
e.g. neglecting $B(r,v)$ completely in the hypernetted chain approximation, 
a first guess of the potential $v^{(1)}$ can be calculated.
With a Monte Carlo (MC) or Molecular Dynamics (MD) simulation the pair 
correlation function $g(r)^{(1)}$ and the direct correlation 
function $c(r)^{(1)}$ for a system of particles interacting via  
$v^{(1)}$ can be determined. 
Thus, the bridge function  $B(r,v^{(1)})$ is determined exactly via 
Eq.~\ref{eq-B}. $B(r,v^{(1)})$ is assumed to be a better approximation for 
$B(r,v)$ than the complete neglect in the first approximation. Inserting
$B(r,v^{(1)})$ in equation \ref{eq-hd} gives the LWR iteration formula:
\begin{eqnarray}
\nonumber &&[v^{(n)}-v^{(n-1)}]/kT=  \mathrm{ln} (g^{(n-1)}/g^{(exp)}) \\&&+ c^{(n-1)} - c^{(exp)} - g^{(n-1)} + g^{(exp)}\label{eq-iter}
\end{eqnarray}

The empirical potential Monte-Carlo (EPMC) scheme \cite{EPMC} 
is closely related to  equation \ref{eq-iter}, 
but considers the upper line of Eq.~\ref{eq-iter}, the logarithmic term only. 
The complete form of Eq. \ref{eq-iter}
has been preferred here, as this scheme shows faster  convergence \cite{RLW}. 
The applicability of the LWR scheme to polyatomic systems has been shown 
by Kahl {\it et al} \cite{kahl}. An example of the application of this technique has been given recently in \cite{phy_b}.

\section{Deduction of the PPDF}
  
In Fig.~\ref{abb-scubr-cubr} the total pair distribution function and the 
total structure factor determined from the hard X-ray experiment are shown.
The  contribution of $g_{CuCu}$ and $g_{BrBr}$ to 
$^xg(r)$ $^{(-CuBr,x)}g(r)$ is also shown in Fig.~\ref{abb-scubr-cubr}. 
$^{(-CuBr,x)}g(r)$ is calculated according to Eq. \ref{eq-conv} 
as the sum of the
PPDF convoluted with ${\rm FT}(w_{ij}(Q))$. It can be seen, that the
first peak in r-space is entirely due to the CuBr-PPDF and, hence,
the height of this peak is independent of possible errors in the
estimation of  the BrBr- and CuCu- PPDF. The corresponding CuBr-PSF is 
shown in the lower part of Fig.~\ref{abb-scubr-cubr}. The height of the 
peak in $g_{CuBr}$ (Fig.~\ref{abb-scubr-cubr}) is 5.2,
clearly higher than than the 3.7 which can be read from Fig.~4 of 
ref.~\onlinecite{All92}. 

Starting point for the determination of the CuCu- and the BrBr-PPDF are the
total neutron pair distribution functions \cite{All92} for $^{63}$CuBr, 
$^{63}$Cu$_{0.5} ^{65}$Cu$_{0.5}$Br and $^{65}$CuBr. 
The CuCu- and the BrBr-PPDF used in the preceding paragraph are the 
result of a direct inversion of the equation system \ref{eq-conv}. 
Two additional constraints to the PPDF have been applied, one 
in Q-space and one in r-space: The CuCu- and 
BrBr-PPDF show only broad features compared to the CuBr-PPDF, the 
corresponding PSF oscillate only up to a momentum 
transfer $Q \sim10$\,\AA$^{-1}$ and are set to zero for $Q>10$\,\AA$^{-1}$. 
A minimum allowed distance has been set in r-space to be 
2.2 \AA\ for the CuCu-PPDF and 3.1 \AA\ for the BrBr-PPDF. The CuCu- and 
BrBr-PPDF determined in this way is more structured 
than the maximum entropy\cite{Ski84} (ME) solution  given 
by Allen {\it et al} \cite{All92}, as it is to be expected. 
Using this ME-solution could only reduce the 
structure in the contribution of these PPDF to the total X-ray pair 
distribution function in Fig.~\ref{abb-scubr-cubr}. 

Fig.~\ref{abb-vgl-tech-cubr} compares the shape of the first CuBr 
coordination shell  determined with EXAFS \cite{DiC97}, neutron
diffraction \cite{All92} and in this work. It is evident, that the
ME solution by Allen {\it et al} \cite{All92}
 underestimates the  sharpness of the CuBr peak. DiCicco {\it et al} 
\cite{DiC97}, who were the first to deduce a sharper CuBr peak, forwarded the 
view that this improved result was due to ''the exceptional short-range 
sensitivity'' of ´the EXAFS experiment. It is in fact the result of the 
use of additional information. Two EXAFS experiments at the bromine 
and copper edge are in principle incapable to extract three PPDF. 
Only with the assignment of the first peak in real space to the 
CuBr-PPDF and the assumption of a Gaussian distribution of the first 
neighbor shell the problem becomes tractable. This assumption is of course 
very reasonable for a system, which melts from a fast ion conductor. 
Given that,
it would be really astonishing, if the like atoms would be the next 
neighbors in this system. 

Introducing the constraints mentioned in the 
preceding paragraph is largely equivalent to the assumption of next 
neighborhood of unlike atoms. Thus diffraction gives the information about 
a sharp CuBr-coordination and, in fact the information about the 
entire pair distribution function. The very good agreement of the CuBr-PPDF
derived form neutron and X-ray diffraction is noted
(Fig.~\ref{abb-vgl-tech-cubr}), if the same constraints are imposed to the 
analysis of the neutron data. The PPDF determined by the maximum entropy 
technique are not the real PPDF, nor do they pretend to be the real PPDF.
 These PPDF have to be interpreted as a lower limit 
estimation of the structure based on no other information  but the neutron
diffraction data. They are, in a sense, a prove that the unlike atoms are 
next neighbors. 

Fig.~\ref{twopeak} shows a close-up of the first CuBr coordination shell.
The peak is strongly asymmetric and clearly cannot be fitted with a single 
Gaussian distribution. It has been fitted with two Gaussians. 
The coordination number integrated over both peaks is four.
\section{Simulation}

In order to interpret the PPDF in terms of a three dimensional 
structure the Levesque, Weis, Reatto potential inversion scheme has 
been applied to the PPDF shown in Fig.~\ref{abb-cubr-rmc-gr}.
It is evident, that the converged LWR-potentials produce a three 
dimensional structure in agreement with these PPDF: According to Pusztai
and McGreevy \cite{Pus98} the same is not true for the 
maximum entropy PPDF of Allen and Howe. Contrary to the hypothesis of Ginoza 
{\it et al}, that the cation-cation PPDF is structureless in all melts of
fast cation conductors, the CuCu PPDF is found similarly structured as the
BrBr-PPDF. This finding is in agreement with the RMC study \cite{Pus98}. 
The agreement of the RMC-PPDF with the PPDF determined here is good for the 
first CuBr coordination shell and the long distance correlations. Some 
discrepancies appear at intermediate distances ($\sim 3-6$\AA). Three facts 
are certain about the PPDF determined here: They are consistent with all 
available diffraction information including the new X-ray result. The set of 
PPDF can be mapped to a three-dimensional arrangement of atoms of the correct 
density. An effective two-body potential exists with the property, that an 
ensemble of particles interacting via this potential samples the 
configuration space consistent with these PPDF.
%

Some of the angular correlations of near neighbor atoms have been 
determined from a MC simulation with the converged LWR-potential. 
These are compared in Fig.~\ref{abb-winkel-cubr} to the 
same functions obtained by RMC \cite{Pus98}. Although there are 
some differences in the underlying PPDF (cf. Fig.~\ref{abb-cubr-rmc-gr}), 
the angular correlations are found to be similar. 
An interesting detail is perhaps, that the maximum in the Br-Cu-Br 
distribution function is shifted form $\cos(\angle_{\rm BrCuBr})\sim -1/3$ 
to $\cos(\angle_{\rm BrCuBr})\sim 0$, 
but it is also evident that quoting the maximum of the distribution 
alone would be misleading.

Fig.~\ref{abb-comp-ang-cubr-zncl2} compares the angular correlations 
between neighboring atoms in ZnCl$_2$ \cite{JPC} and CuBr, both obtained using 
the LWR formalism. Both melts are based on a tetrahedral structural 
motive and in fact Cu(I) and Zn(II) are isoelectronic species. So some of 
the distribution functions, the $-+-$ and the $---$ distribution function, 
show similarities. However, there is a pronounced difference in the 
distribution function of the $+-+$ angle, 
the angle joining adjacent tetrahedra. While in 
ZnCl$_2$ there is a peak at $\cos(\angle_{+-+})\sim -1./3.$
the corresponding angular correlations in CuBr are very weak. As a consequence also the $+++$ correlation is much less structured 
in CuBr than in ZnCl$_2$.
Thus a continuous
random network is a useful parameterization for the ZnCl$_2$, but not the 
CuBr structure. The fact, that ZnCl$_2$ but not CuBr can be easily
supercooled into a glass is probably intimately related to this difference.

\section{Conclusion}
The short range order of the melt of the fast-ion conductor 
Cu(I)Br has been discussed recently. By diffraction experiments 
with hard electromagnetic radiation it can be conclusively shown, that the 
first CuBr coordination shell is definitely sharper than obtained earlier 
in a maximum entropy analysis of NDIS data by Allen {\it et al}. 
This is in agreement with the 
result obtained by DiCicco {\it et al} using EXAFS spectroscopy. 
The analysis presented in this work indicates, that the 
featureless CuCu-PPDF obtained by Allen {\it et al} is an artifact of the 
maximum entropy analysis, contrary to the view that a 
structureless cation-cation PPDF is a common feature in all fast-ion 
conductors. The same result has been obtained by reverse 
Monte Carlo by Pusztai {\it et al}.
 
The LWR potential inversion scheme has been used
to derive an effective two-body potential from the PPDF. This method can 
be used to interpret the PPDF in terms of a three dimensional structure.
A MC simulation with the converged potential has been used to determine 
angular correlations between near neighbor atoms.
The angular correlations in ZnCl$_2$ melt shows
similarities in the $-+-$ distribution functions reflecting the similarity in 
the local environment of the cation. However,
while in ZnCl$_2$ the angle between adjacent ZnCl$_4$ tetrahedra is 
--- like in silica glass --- well defined and the $+-+$ angle distribution 
function is peaked, the corresponding function in CuBr is, as the $+++$ 
function, broad and featureless. This behavior is probably the key to 
understand, why ZnCl$_2$ but not CuBr can be easily supercooled into a 
glassy state.

\begin{figure}
\caption{Raw data of the high energy electromagnetic 
radiation diffraction experiment on CuBr melt at 803\,K (scattering 
contribution from the container, the sample in the container, and the 
sample scattering after subtraction of the container scattering)}\label{abb-roh-cubr}
\end{figure}
\begin{figure}[tbhp]

\caption{The total X-ray PDF, the CuBr PPDF and the
CuBr-PSF of liquid CuBr.\\ Upper part of the figure: The total 
X-ray PDF $^x g(r)$ (shifted -2.0 units in y-direction), 
the contribution of $g_{BrBr}$ and $g_{CuCu}$ to it (see text) and the
resulting CuBr PPDF.\\ Lower part of the figure:
 The  $Q\cdot (S_{CuBr}(Q)-1)$ corresponding to the PPDF in the upper part of the  figure. The symbols are the experimental points, the solid line is 
 the Fourier back-transform of the PDF in the upper part of the figure, the
 dashed line is the neutron derived  $S_{CuBr}$ (see text). Within 
 the momentum transfer range covered by both experiments the solid and 
 dashed line are almost indistinguishable.}\label{abb-scubr-cubr}
\end{figure}
\begin{figure}
\caption{
Comparison of the CuBr-PPDF obtained with various techniques}
\label{abb-vgl-tech-cubr}
\end{figure}
\begin{figure}
\caption{Close-up of the first peak in $g_{\rm CuCu}$. and the Gaussian 
distributions fitted to it. The midpoints, $\sigma$ parameter and 
coordination numbers are given in the figure. The residue is is shown 
shifted by -0.5 units in y-direction.}\label{twopeak}
\end{figure}
\begin{figure}
\caption{PPDF of CuBr(l).\\
Comparison of the experimental PPDF (symbols, cf. preceding section), with 
the PPDF of a MC-simulation with the effective pair-potentials of the 
LWR-scheme (solid line), the RMC-model i) (dashed dotted line) and ii)
(dashed line) from ref. \cite{Pus98} \\
RMC-model i) and ii) differ significantly only for g$_{\rm CuCu}$ 
and thus only $g_{\rm CuCu}$ is shown for model i). Model i) and ii)
have been obtained by applying the RMC formalism to different scattering 
functions, 
either the total structure function of the PPDF.}\label{abb-cubr-rmc-gr}
\end{figure}
\begin{figure}
\caption{Comparison of the angular correlations 
between adjacent atoms in CuBr(l) determined by  RMC-simulation \cite{Pus98} 
and the LWR-scheme.\\
Symbols refer to the RMC simulation, the full lines to the LWR scheme using 
the same cut-off radii as in the RMC work (i.e. Br-Br: 5.2\,\AA, 
Cu-Br: 3.3\,\AA\, CuCu: 5.2\,\AA ) The dashed lines correspond 
to a cut-off radius of  3.3\, \AA\ (CuCu) and 2.8\, \AA\ (CuBr). \label{abb-winkel-cubr}}
\end{figure}
\begin{figure}
\caption{ Comparison of the angular correlations
between neighboring atoms in CuBr(l) (solid line) and ZnCl$_2$(l) (symbols)
\cite{JPC}}\label{abb-comp-ang-cubr-zncl2}
\end{figure}

\begin{thebibliography}{99}
\bibitem{PM71} D.~I.~Page and K.~Mika, J.~Phys.~C: Sol.~State Phys. {\bf 4} 
3034 (1971) 
\bibitem{eisen} S.~Eisenberg, J.-J.~Jal, J.~Dupuy, P.~Chieux and W.~Knoll: 
Phil.~Mag. B {\bf 46} 195 (1982)
\bibitem {Gin87} M.~Ginoza, J.~H.~Nixon and M.~Silbert: J.~Phys.~C 
{\bf 20} 1005 (1987)
\bibitem{All92} D.~A.~Allen and R.~A.~Howe, J.~Phys.~Condens. Matter
Condens. Matter {\bf 4}, 6029 (1992)
\bibitem{Koe91} L. K\"oster, H. Rauch and E. Seymann: 
At.~Data Nucl.~Data Tab. {\bf 49} 65 (1991)
\bibitem{DiC97} A.~DiCicco, M.~Minicucci and A.~Filipponi: Phys.~Rev.~Lett. {\bf 78} 460 (1997)
\bibitem{Pus98} L.~Pusztai and R.~L.~McGreevy: J.~Phys. Condens.~Matter
{\bf 10}  525 (1998)
\bibitem{ego0}  J.~Neuefeind, H.~F. Poulsen: Physica Scripta
{\bf T57}  112 (1995)
\bibitem{LWR} D.~Levesque, J.~J.~Weis and L.~Reatto,  Phys.~Rev.~Lett.
{\bf 54} 451 (1985)
\bibitem{Bou98} R.~Bouchard, D.~Hupfeld, T.~Lippmann, J.~Neuefeind,
H.-B.~Neumann, H.~F.~Poulsen, U.~R\"utt, J.~R.~Schneider, J.~S\"ussenbach and
 M.~v.~Zimmermann: J.~Synchrotron Rad. {\bf 5} 90 (1998)
\bibitem{YB} K.~Ibel(ed.),  {\it Guide to Neutron Research
Facilities at the ILL} (ILL: Grenoble: 1994)
\bibitem{Neu98} J.~Neuefeind, K.~T\"{o}dheide, A.~Lemke and 
H.~Bertagnolli: J.~Non-Cryst.~Sol. {\bf 224}  205 (1998)
\bibitem{MolPhys} T.~Weitkamp, J.~Neuefeind, H.~E.~Fischer, 
M.~D.~Zeidler, Molec.~Phys., {\bf 98}, 125 (2000)
\bibitem{Hub75} J.~H.~Hubell, W.~J.~Veigele, E.~A.~Briggs, R.~T.~Brown,
D.~T.~Cromer and R.~J.~Howerton: J.~Phys.~Chem.~Ref.~Data {\bf 4} 471 (1975)
 \bibitem {Sai97} M.~Saito, C.~Park, K.~Omote, K.~Sugiyama, Y.~Waseda:
J.~Phys.~Soc.~Jpn. {\bf 66}  633 (1997)
\bibitem{EPMC} A.~K.~Soper: Chem.~Phys. {\bf 202}  295 (1996)
\bibitem{RLW} L.~Reatto, D.~Levesque and J.~J.~Weis: 
Phys.~Rev.~A {\bf 33} 3451 (1986) 
\bibitem{kahl} G. Kahl and K. Kristufek: Phys.~Rev.~E,  {\bf 49} 3565 (1994) 
\bibitem{phy_b}  J.~Neuefeind, H.~E.~Fischer and  W.~Schr\"{o}er: 
Physica B,  {\bf  276-278} 481 (2000)
\bibitem{Ski84} J.~Skilling and R.~K.~Bryan: Mon.~Not.~R.~Astron.~Soc.
{\bf 211}  111 (1984) 
\bibitem{JPC} J. Neuefeind: J.~Phys.~Condens.~Matter: submitted (2000)

\end{thebibliography}
\end{document}